\documentclass[manuscript]{acmart}

\AtBeginDocument{%
  \providecommand\BibTeX{{%
    \normalfont B\kern-0.5em{\scshape i\kern-0.25em b}\kern-0.8em\TeX}}}

\copyrightyear{2026}
\acmYear{2026}
\usepackage{multirow}
\usepackage{amsmath}
\usepackage{mathtools}
\usepackage{graphicx}
\usepackage{listings}
\usepackage{xcolor}
\usepackage{verbatim}
\usepackage{minted}
\usepackage{tcolorbox}
\tcbuselibrary{breakable}
\usepackage[utf8]{inputenc}
\usepackage{xcolor}
\usepackage{hyperref}
\usepackage{enumitem}
\usepackage{tikz}
\usepackage{graphicx}
\usepackage{tabularx}
\usepackage{booktabs}

\begin{document}

\title{From Logs to Agents: Reconstructing High-Level Creative Workflows from Low-Level Raw System Traces}

\author{Tae Hee Jo}
\email{persontae94@gmail.com}
\affiliation{%
  \institution{Design Informatics Lab, Hanyang University}
  \city{Seoul}
  \postcode{04763}
  \country{Republic of Korea}
}
\affiliation{%
  \institution{Human-Centered AI Design Institute, Hanyang University}
  \city{Seoul}
  \country{Republic of Korea}
}

\author{Kyung Hoon Hyun}
\authornote{Corresponding author.}
\email{hoonhello@gmail.com}
\affiliation{%
  \institution{Design Informatics Lab, Hanyang University}
  \city{Seoul}
  \postcode{04763}
  \country{Republic of Korea}
}
\affiliation{%
  \institution{Human-Centered AI Design Institute, Hanyang University}
  \city{Seoul}
  \country{Republic of Korea}
}

\renewcommand{\shortauthors}{Tae Hee Jo and Kyung Hoon Hyun}

\begin{abstract}
Current AI-based Creativity Support Tools (CSTs) generate massive amounts of low-level log data (e.g., clicks, parameter tweaks, metadata updates) that are hard to interpret as "creative intent". We argue that to enable future agentic systems to understand and assist users, we must first translate these noisy system traces into meaningful high-level user behavioral traces. We propose a method that parses raw csv/JSON logs into structured behavioral workflow graphs that map the provenance and flow of creative assets. By abstracting low-level system events into high-level behavioral tokens (e.g., \texttt{MODIFY\_Prompt}, \texttt{GENERATE\_Image}), this method enables downstream analyses like sequence mining and probabilistic modeling. We discuss how this structured workflow history is a prerequisite for 'Process-Aware Agents'—systems capable of suggesting next design moves or explaining rationales based on a deeper understanding of the user's workflow patterns and history.
\end{abstract}

\keywords{creative activity traces, creativity support tools, design process analysis, workflow, agent enhancement}

\maketitle

\section{Introduction}
The landscape of Creativity Support Tools (CSTs) has shifted from linear editing software to complex, node-based generative environments such as ComfyUI \cite{comfy_org}, Flora AI \cite{flora_ai} and GENPRESSO \cite{genpresso} While these tools offer high creativity, they present a significant challenge for both researchers and system designers: the "black box" of the creative process. While we possess rich raw data in the form of csv/JSON system logs, and theories of creativity such as divergent versus convergent thinking \cite{lee2024impact, goldschmidt2014linkography}, we lack a computational representation of the user's creative journey that is both machine-readable and semantically meaningful.

The gap is technical and semantic. Raw logs from node-based CSTs can be extremely noisy. In our analysis of generative design sessions, we observed that a single "creative design move" such as tweaking a prompt to change a lighting effect, can generate dozens of system events, including connection re-routing, system cleanup routines, and redundant metadata updates. Analyzing creativity at this resolution is inefficient, as it is both computationally difficult and cognitively loading \cite{he2022empirical}.

We propose a workflow reconstruction pipeline that automates the translation of low-level system trace logs into high-level user behavioral workflow graphs. We argue that this restructuring data process is necessary to enhance the next generation of CST agents. Current agents operate on state (the current canvas); to be truly collaborative, they must understand the full workflow process history (how the user arrived there).

\section{The Proposed Method}
We present a three-stage pipeline designed to transform raw csv/JSON logs into actionable user behavioral insights. This method has been applied and tested on generative design logs from a node-based CST (GENPRESSO).

\subsection{Semantic Filtering and Classification (De-noising)}
The first challenge in analyzing raw trace data is distinguishing explicit high-level user intent from low-level system noise. Raw data logs often contain echoes of user actions which are redundant internal server calls, state synchronization events, and temporary object lifecycle markers that do not reflect a new creative decision. 

We implement a heuristic classification layer that filters events based on \texttt{Action\_Type} and the system's \texttt{Raw\_Source\_Label} to isolate meaningful creative moves.
\begin{itemize}
\item \textbf{Filtering System Noise:} Our pipeline automatically discards events that represent system-side maintenance rather than user intent.  This includes cleanup events where the system purges temporary node or asset IDs, purely structural updates related to node connection logic (e.g., backend graph re-routing), and intermediate state updates that occur during asynchronous generation processes. By removing these artifacts, we prevent the analysis from interpreting system latency or housekeeping as user hesitation or activity.
\item \textbf{Heuristic Classification:} We classify the remaining high-intent events into four creative design moves based on a rule-based mapping of the raw logs. \textbf{INSERT:} Represents the introduction of new content into the canvas space. This includes importing external images, creating new nodes; \textbf{MODIFY:} Captures the refinement of existing assets. This includes altering context within a prompt, image, or video nodes, or adjusting the generation output parameters (e.g., AI model, aspect ratio); \textbf{GENERATION:} Distinct from simple insertions, this classification is reserved for the execution of the generative AI model itself, marking the moment a user commits their parameters to produce a new artifact (image or video); \textbf{REMOVE:} Represents the explicit deletion of assets, distinguishing between a user rejecting a generated outcome and the system clearing a temporary cache.
\end{itemize}

In our pilot dataset, this filtering layer reduced the event volume by approximately 40\% (from 927 raw system logs to 563), successfully isolating sequences that represent tangible changes to the creative artifact while discarding backend redundancy.

\subsection{Design Sequence Reconstruction (Visualizing the workflow history)}
Creativity in generative tools is rarely linear \cite{smith2025fuzzy}; it is branching. A user may generate an image, branch off to try a variation, abandon it, and return to the parent node. Timestamps alone are insufficient to track this evolution within the user's creative design process, as they only record \textit{when} an action happened, not \textit{where} it fits in the design ideation hierarchy.

\begin{figure}[h]
  \centering
  \includegraphics[width=\linewidth]{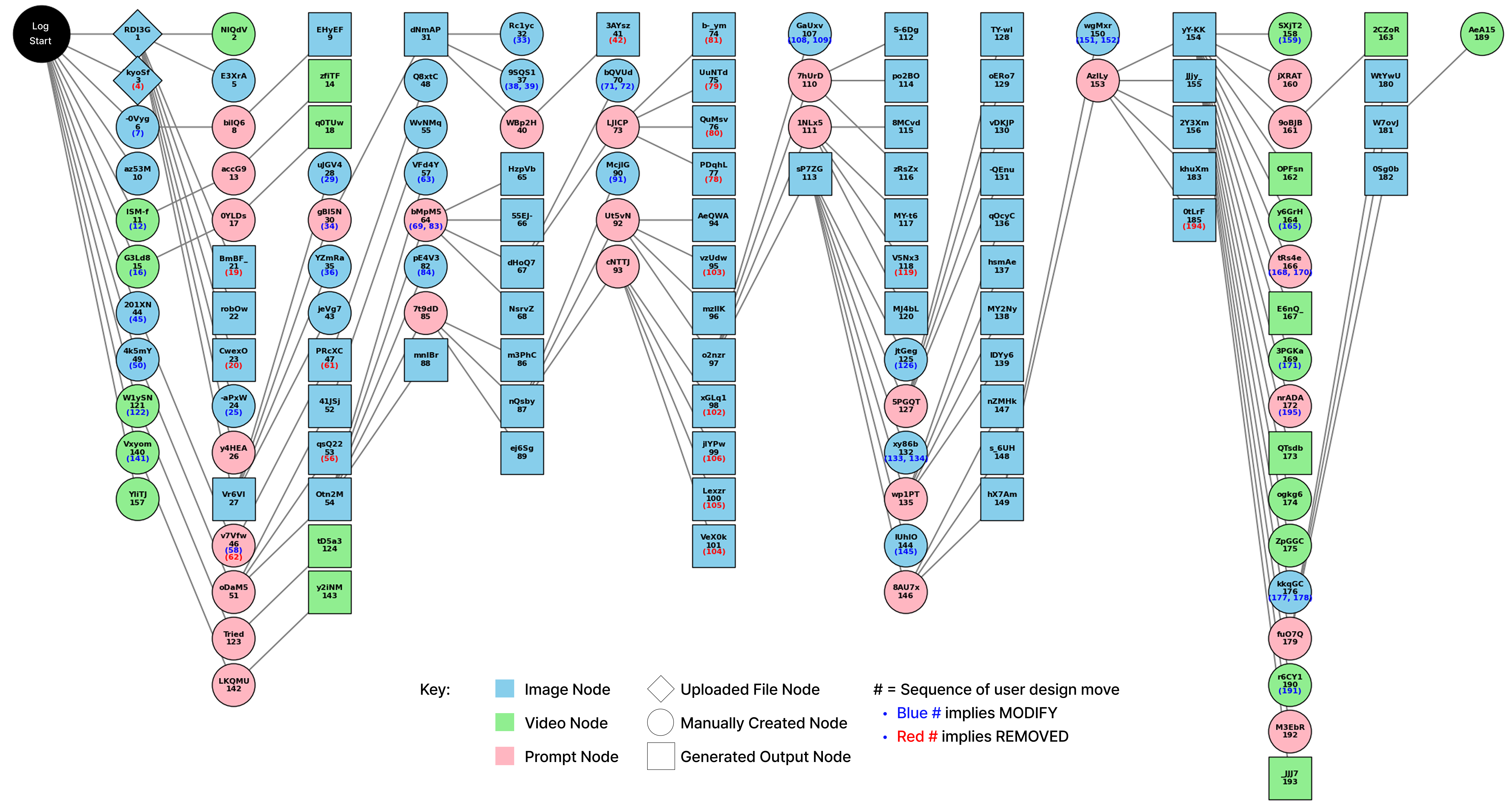}
  \caption{This abstract user behavioral workflow graph visualizes the branching evolution of a creative design process session. Blue nodes indicate image nodes, green nodes indicate video nodes, and pink nodes indicate prompt nodes. The structure reveals a user exploring multiple parallel variations before converging on a single lineage.}
  \Description{A directed acyclic graph showing nodes connected by arrows. The graph flows from left to right, showing branching paths where a single node splits into multiple outcomes.}
  \label{Figure1}
\end{figure}

We construct a Directed Acyclic Graph (DAG) (\ref{Figure1}) by resolving parent-child relationships extracted from the raw metadata source fields (e.g., \texttt{connected\_from} fields). The X-axis represents the generation depth (distance from the log start) and Y-axis represents the chronological order of log events specific to its generation depth. Each node contains unique number labels indicating the sequence number of each user's creative design move; black number indicates the global sequence order, blue number indicates the sequence of the MODIFY design move, and red number indicates the sequence of the REMOVE design move. Nodes represent creative assets (Prompt Nodes, Image Processes, Video Outputs) and edges represent the flow of data. Square nodes indicate an AI generated output, while circle nodes indicate a user's manually created creative asset. 

We utilize a depth-based layout algorithm (implemented in Python/NetworkX) to visualize the creative design session. As shown in Figure 1, this reveals the overall shape of the design process workflow of the user's CST design session, distinguishing between wide exploration (many branches) and deep refinement (long branches).

\subsection{Tokenization}
To make these graphs analyzable, we convert the nodes into standardized tokens. We concatenate the user action and node asset type to create a standardized language of interaction such as, \texttt{INSERT\_prompt}, \texttt{MODIFY\_image}, \texttt{GENERATION\_video}. This tokenization abstracts away the specific tool interface, allowing for potential cross-platform comparison.

\section{Implications: Enhancing Process-Aware Agents}
While the primary contribution of this work is the reconstruction pipeline of low-level system traces to high-level user behavioral workflow itself, its value lies in the downstream applications for Agentic AI for CSTs.

\subsection{From Descriptive to Predictive Analysis}
By converting logs into behavioral tokens, we can move from describing what happened to modeling the probability of what happens next. In our pilot sequence analysis of 563 log events, we identified distinct probabilistic signatures. For instance, The most common Bigram (2-step sequence) was \texttt{GENERATION\_image} - \texttt{GENERATION\_image} (19.1\%, Frequency=37). This indicates a high frequency of 'blind' re-rolling, suggesting the user is relying on the stochastic nature of the model rather than refining their input. 

To further quantify these behaviors, we applied Markov Chain analysis to model state transition probabilities \cite{thimbleby2001usability}. We found that after a user inserts an image node (\texttt{INSERT\_image}), there is a 69.6\% probability that their immediate next action is to modify its content (\texttt{MODIFY\_image}). Conversely, after generating an image, there is a 66.1\% probability of immediate re-generation. These transition probabilities provide a mathematical basis for an agent to distinguish between "setup phases" (high probability of modification) and "exploration phases" (high probability of re-generation).

\subsection{Scaffolding Agentic Decisions}
We argue that the user behavioral workflow history graph data should be fed into the context window of future CST agents (e.g., LLM-driven assistants), as an agent cannot effectively collaborate if it has no memory of the workflow history. For instance, if a user has performed the sequence \texttt{INSERT\_image} - \texttt{MODIFY\_metadata\_update} (resizing image) five times in a row, current agents would wait for the user's next command. However, a process-aware agent, aware of the high transition probability (69.6\%) between insertion and modification, could then infer the intent and offer \textit{"I see you are importing and resizing references. Shall I automatically apply this scale factor to future imports?"} Therefore, by analyzing the workflow history graph data, an agent can provide better rationale. Instead of saying \textit{"Try this prompt,"} it could suggest smarter decision making moves \textit{"Users who followed a similar generation path found success by \{pattern context\}."}

\section{Conclusion}
We proposed a workflow reconstruction pipeline to turn low-level raw system data into high-level, actionable behavioral insights. By filtering noise, classifying raw data, reconstructing design sequence lineage, and tokenizing actions, we provide a structured foundation for analyzing user behavioral design process workflows based off raw log data in node-based CSTs.
Beyond analysis, this structured data serves as a critical enabler for the next generation of AI agents. We argue that by equipping agents with a "memory" of the user's behavioral design process history, we can transition from passive tools that merely execute commands to process-aware collaborators. Such agents could leverage historical workflow patterns to predict user intent, suggest strategic design moves, and provide rationale grounded in the provenance of the creative process.

\section*{Acknowledgments}
This work was supported by the Industrial Technology Innovation Program(RS-2025-02317326, Development of AI-Driven Design Generation Technology Based on Designer Intent) funded by the Ministry of Trade, Industry \& Energy (MOTIE, Korea).

\bibliographystyle{ACM-Reference-Format}
\bibliography{references}

\end{document}